\documentclass[conference]{IEEEtran}
\IEEEoverridecommandlockouts
\usepackage{cite}
\usepackage{amsmath,amssymb,amsfonts}
\usepackage{algorithmic}
\usepackage{graphicx}
\usepackage{textcomp}
\usepackage{xcolor}
\usepackage{enumitem}
\usepackage{hyperref}

\def\BibTeX{{\rm B\kern-.05em{\sc i\kern-.025em b}\kern-.08em
    T\kern-.1667em\lower.7ex\hbox{E}\kern-.125emX}}
\begin{document}

\title{\textbf{EYWA}: \textbf{E}lastic load-balancing \& high-availabilit\textbf{Y} \textbf{W}ired virtual network \textbf{A}rchitecture}

\author{\IEEEauthorblockN{1\textsuperscript{st} Wookjae Jeong}
\IEEEauthorblockA{\textit{Dept. of Distributed Architecture} \\
\textit{Independent Researcher}\\
Seoul, Korea \\
wjjung11@gmail.com}
\and
\IEEEauthorblockN{2\textsuperscript{nd} Jungin Jung}
\IEEEauthorblockA{\textit{Dept. of Distributed Architecture} \\
\textit{Independent Researcher}\\
Seoul, Korea \\
call518@gmail.com}
}

\maketitle

\begin{abstract}
Infrastructure as a Service (IaaS) in cloud environments provides compute, storage, networking, and other fundamental resources that allow consumers to deploy and run arbitrary software, including operating systems and applications. To support multi-tenant environments, IaaS leverages virtualization, but conventional overlay network architectures have become a direct cause of scalability limitations. In particular, current IaaS virtual networks face challenges in high availability and load balancing.

To address these issues, we present EYWA, a virtual network architecture that scales to support very large data centers with high availability, efficient load balancing, and large layer-2 semantics. EYWA overcomes scalability limitations by: (1) accommodating a large number of tenants (about $2^{24}$ = 16,777,216) through logically isolated virtual LANs with unique IP ranges, (2) providing per-tenant public network services without throughput bottlenecks or single points of failure in network address translation (SNAT/DNAT), and (3) enabling a single large IP subnet per tenant with extended layer-2 semantics.

EYWA combines existing techniques into a distributed scale-out control and data plane. Its only component is an agent running on each hypervisor host, which collectively act as a distributed controller. As a result, EYWA can be deployed in today’s multi-tenant cloud environments. We have implemented a proof-of-concept (PoC) of EYWA and evaluated its effectiveness through measurements and experiments in our lab.
\end{abstract}

\bigskip

\begin{IEEEkeywords}
Cloud, IaaS, Virtual network, Overlay network, Data center network, OpenStack, Load balancing, Load sharing
\end{IEEEkeywords}

\bigskip

\section{\textbf{Introduction}}
After the emergence of the new paradigm of cloud computing, different forms of IaaS services have appeared. First, some service providers \cite{ref20}, \cite{ref22}, \cite{ref25}, \cite{ref26} offer public cloud services using their in-house solutions. Second, there are solution vendors \cite{ref21}, \cite{ref23} that provide proprietary software, which service providers then use to deliver public or private cloud services. Third, project groups \cite{ref16}, \cite{ref17}, \cite{ref18}, \cite{ref19} develop open-source platforms, and other providers \cite{ref24}, \cite{ref27} deliver public or private cloud services based on them.

While these efforts have addressed and improved many technical limitations of IaaS, network virtualization in cloud environments still faces numerous challenges and has progressed slowly. In contrast, physical (underlay) data center network architectures for cloud environments have been extensively researched \cite{ref1}, \cite{ref2}, \cite{ref3}, \cite{ref8}, \cite{ref9}] and standardized \cite{ref13}, \cite{ref14} to overcome the limitations of the conventional three-tier model and to identify optimal network designs. Unfortunately, comparable advances in virtualized networking remain limited.

Each existing open-source solution introduces its own network models, which are sometimes similar, and in some cases, multiple models coexist within a single solution. This indicates that an optimal model has yet to be established. Proprietary in-house solutions used by providers remain black boxes, but based on publicly available information, they also suffer from similar issues because they rely on conventional architectures, protocols, and infrastructures.

\begin{figure}[htbp]
    \centering
    \fbox{\includegraphics[width=0.9\linewidth]{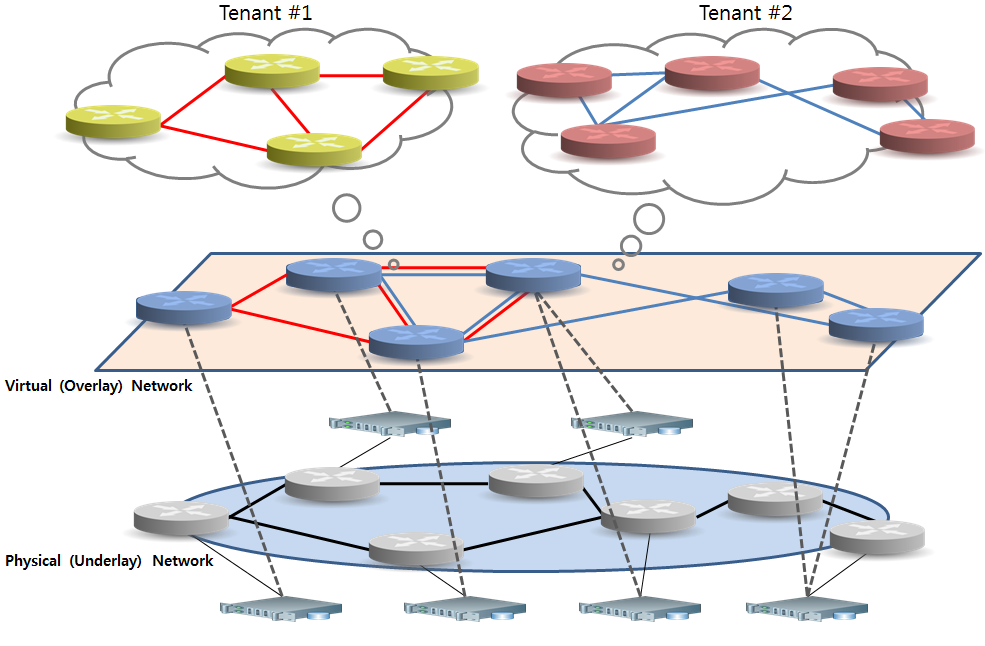}}
    \caption{Virtual (Overlay) network for multi-tenant cloud}
    \label{figure-1}
\end{figure}

\begin{figure}[htbp]
    \centering
    \fbox{\includegraphics[width=0.9\linewidth]{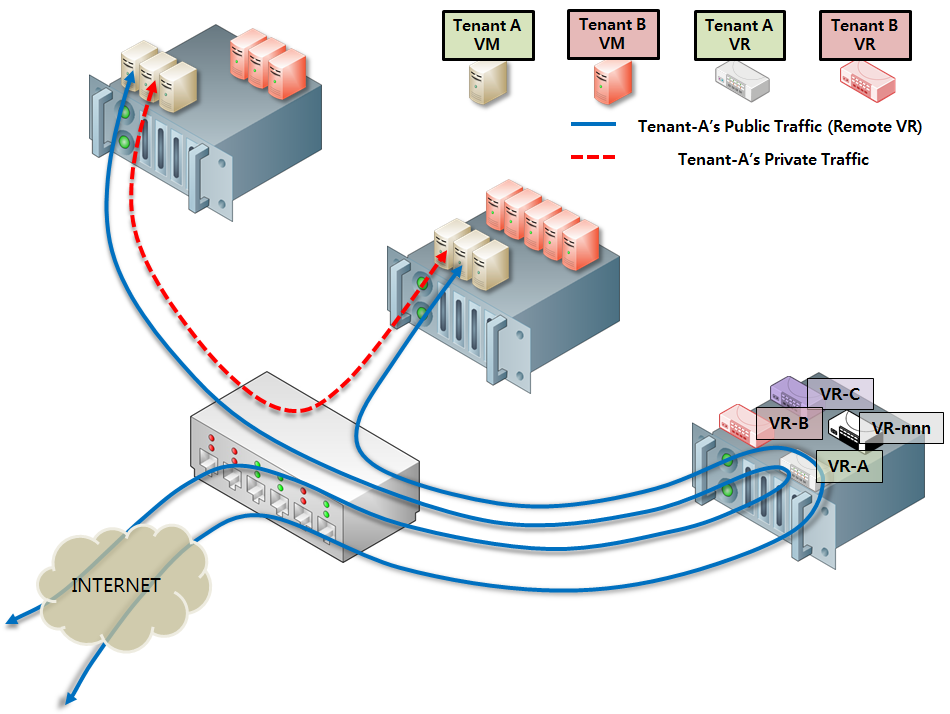}}
    \caption{Traffic Flows in a shared network service host}
    \label{figure-2}
\end{figure}

\bigskip

\section{\textbf{BACKGOUND}}
\label{sec:background}

In this section, we describe the dominant issues in today’s virtual network architectures. When a user creates VMs in multi-tenant cloud environments, each tenant is typically assigned a virtual LAN for private networking, a Virtual Router (VR) with a public IP address for external connectivity, and private IP addresses for each VM. The VR may also be assigned additional public IP addresses on behalf of its VMs. Specifically, a VM’s private IP is directly allocated to its virtual network interface, while the tenant’s public IP is indirectly assigned to the VR, which can be implemented as a VM instance or a Linux network namespace. The VM then communicates with external networks through the VR’s SNAT/DNAT functions. Outbound traffic uses the VR’s SNAT function as the default gateway, and inbound traffic through the public IP is handled via DNAT. 

\begin{figure}[htbp]
    \centering
    \fbox{\includegraphics[width=0.9\linewidth]{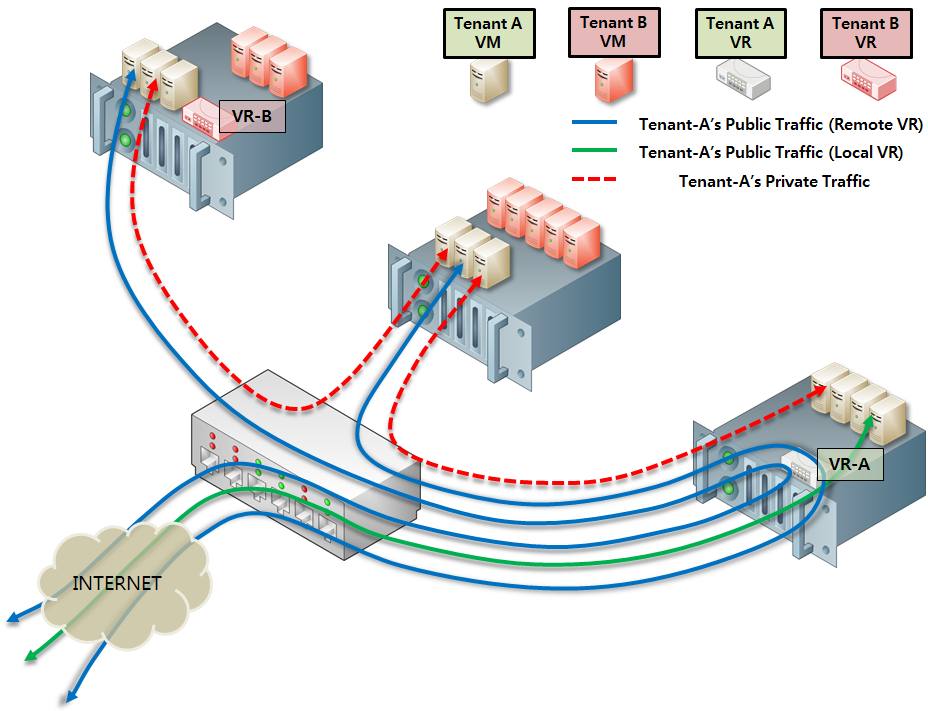}}
    \caption{Traffic flows in hypervisor network service hosts}
    \label{figure-3}
\end{figure}

Figure \ref{figure-5} [16, 18] and Figure \ref{figure-3} \cite{ref17} illustrate how public traffic flows through a VR in multi-tenant environments, using either a shared network service host model or a hypervisor-based service host model. Inter-VM communication within the same virtual LAN (e.g., via VLAN \cite{ref11}, VxLAN \cite{ref5}, NVGRE \cite{ref6}, or STT \cite{ref7}) may occur directly over layer-2 protocols, or indirectly through a centralized network service host.

Although this architecture may appear complex, it essentially mirrors conventional physical network designs in virtualized environments. Consequently, virtual network architectures inherit the same scalability challenges found in physical networks, particularly regarding high availability and load balancing.

\bigskip

The conventional approach has the following problems:

\bigskip

\textbf{Public Network} has the following problems because VMs rely on the SNAT/DNAT function of a single VR.

\begin{itemize}
  \item \textbf{High Availability}: In a shared network service host model (e.g., OpenStack/Network Node), a single point of failure (SPOF) exists, as illustrated in Figure \ref{figure-2}. In the worst case, if the shared service host fails, all VMs in the affected cluster (not only a single tenant) lose external connectivity. By comparison, in a hypervisor-based service model, the failure domain is limited to a single tenant (Figure \ref{figure-3}); in the worst case, only the VMs of that tenant lose connectivity if the host running the VR fails. Protocols such as VRRP \cite{ref12} have been proposed to enhance availability, but even Active-Standby or limited Active-Active configurations with additional VRs or service hosts cannot fully ensure high availability.
  
  \item \textbf{Load Balancing}: A single VR or shared network service host inevitably becomes a throughput bottleneck for SNAT/DNAT processing and layer-4 load balancing. As a result, the total number of VMs per virtual LAN is constrained by the performance and bandwidth of that single VR or host, even if the infrastructure could support more VMs. For example, scaling out public services such as web servers is hindered by VR capacity. Some IaaS providers \cite{ref27} employ additional physical load balancers (scale-up), while others offer dedicated services such as AWS Elastic Load Balancing (ELB) \cite{ref30}. However, none of these approaches achieve unlimited scalability.
  
  \item \textbf{Traffic Engineering}: In a hypervisor service model, most public traffic must traverse a remote VR, as illustrated by the blue line in Figure \ref{figure-3}. This consumes additional bandwidth and increases latency. Only traffic from VMs sharing the same host as the VR (green line in Figure \ref{figure-3}) avoids this overhead. In the shared service model, all public traffic is forced through the remote service host (Figure \ref{figure-2}), further amplifying bandwidth consumption and latency.
\end{itemize}

\bigskip

\textbf{Private Network} is required because cloud tenants demand that their VMs reside in different layer-2 subnets or layer-3 networks from others, to ensure security and traffic isolation in multi-tenant environments. To achieve this, most platforms provide a dedicated virtual LAN per tenant, typically implemented with VLAN (802.1Q) \cite{ref11}.

\begin{itemize}
  \item \textbf{VLAN (802.1Q) limit}: A cloud data center can quickly exhaust the VLAN ID space of 4,094, especially with many top-of-rack switches connected to multiple physical servers hosting VMs that each require at least one VLAN. This limitation hinders tenant expansion, restricts VM communication across layer-2 domains, and complicates VM mobility. Workarounds exist, but they inevitably increase the complexity of the network architecture.
  
  \item \textbf{A single large IP subnet (large layer-2 network)}: To fully leverage layer-2 communication, a large number of VMs can be deployed in a single virtual LAN. However, this approach introduces additional challenges such as Address Resolution Protocol (ARP) broadcast overhead, MAC table flooding, and Spanning Tree Protocol (STP) issues, all of which must be resolved to scale effectively.
\end{itemize}

\bigskip

In this paper, we first propose a virtual network architecture that improves scalability using only conventional protocols, then discuss its remaining limitations. We then introduce EYWA, a design that achieves virtually unlimited scalability to further overcome these challenges.

\bigskip

\section{\textbf{Limited Scalability}}

This section reviews a virtual network architecture that uses only conventional protocols to mitigate existing problems. In this design, the VR integrates SNAT/DNAT and layer-4 load balancing functions into a single VM instance, following the hypervisor network service model illustrated in Figure \ref{figure-3}. 

In addition, the design assumes the use of external DNS-based load balancing, such as AWS Route 53 \cite{ref31}, combined with services like AWS Elastic Load Balancing (ELB) \cite{ref30}, to complement the VR’s built-in layer-4 load balancer for public services.

\bigskip

\begin{figure}[htbp]
    \centering
    \fbox{\includegraphics[width=0.9\linewidth]{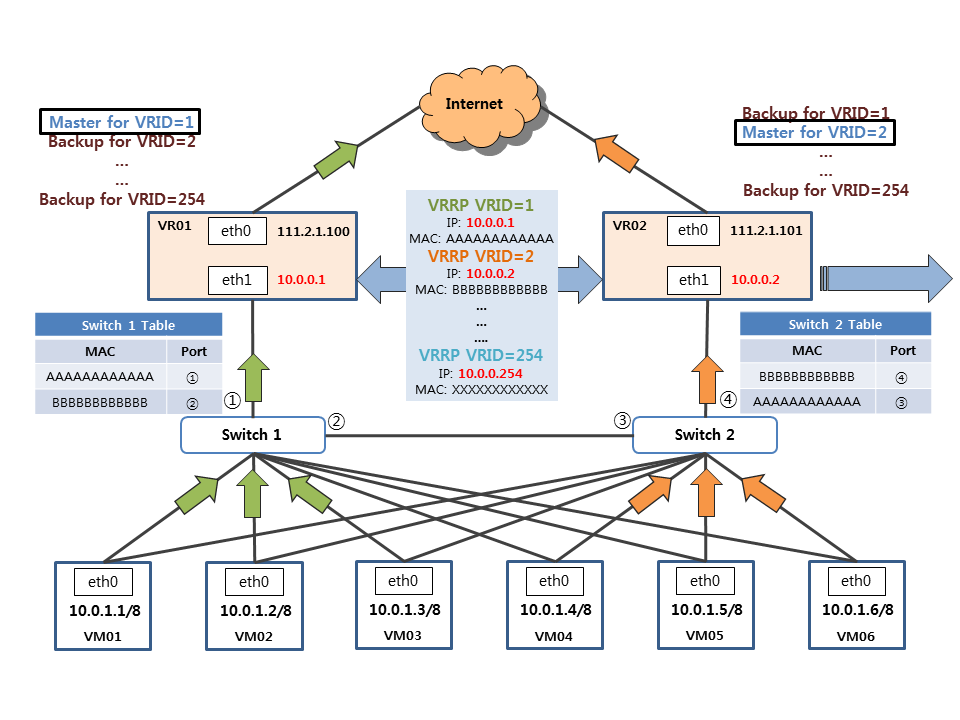}}
    \caption{SNAT traffic flows with multiple VRRP groups}
    \label{figure-4}
\end{figure}

\textbf{Public Network} challenges can be addressed by applying the Multiple Virtual Router Redundancy Protocol (MVRRP \cite{ref12}), as illustrated in Figure \ref{figure-4}. MVRRP supports up to 255 VRRP groups per network interface, enabling each VR to implement redundancy and load sharing within a virtual LAN topology.

\begin{figure}[htbp]
    \centering
    \fbox{\includegraphics[width=0.9\linewidth]{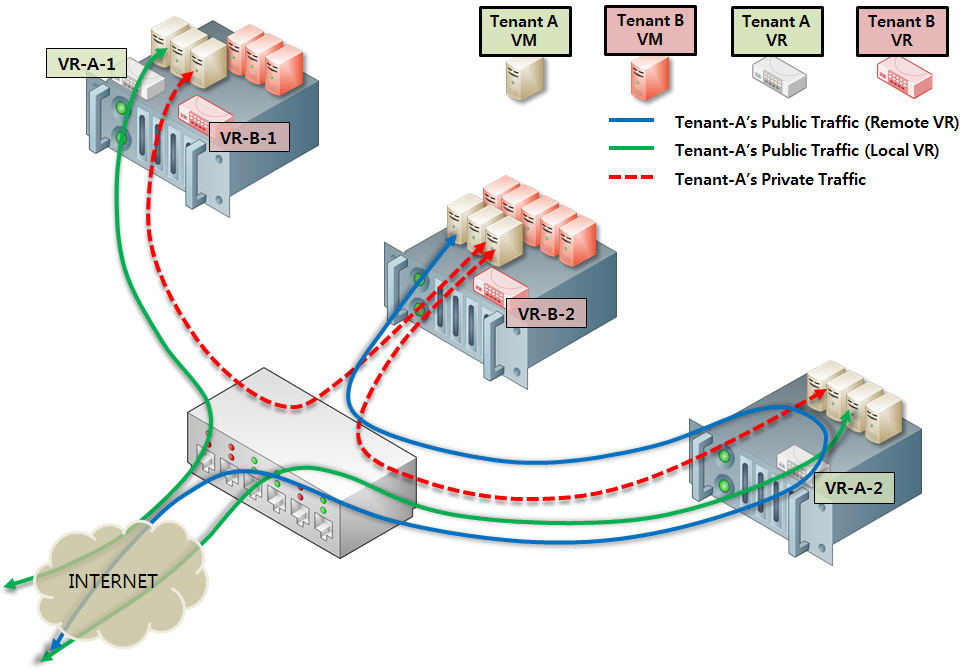}}
    \caption{Traffic flows on MVRRP architecture}
    \label{figure-5}
\end{figure}

\begin{itemize}
  \item \textbf{High Availability}: An active-active configuration with multiple VRs (up to 254 per tenant, within the VRID limit) prevents SPOFs and ensures high availability. Each VR is assigned a distinct private IP address (e.g., 10.0.0.1–254) and can scale out to as many as 254 instances. When a VM is instantiated, if a local VR exists, its private IP is assigned as the VM’s default gateway. Otherwise, a suitable remote VR among the available instances is selected according to policies such as round-robin, latency, or performance. The VM’s public IP is then mapped to that default gateway VR. If a master VR fails, a backup VR automatically takes over both the private IP (as the VM’s gateway) and the associated public IPs. This provides seamless high availability.
  
  \item \textbf{Load Sharing and Balancing}: VM default gateways are distributed across up to 254 VRs according to policy when instantiated, thereby reducing outbound throughput bottlenecks caused by a single VR. Inbound traffic is also balanced using per-VR software load balancers in combination with external DNS-based load balancing.
  
  \item \textbf{Traffic Engineering}: VMs with a local VR benefit from reduced bandwidth consumption and lower latency, since their traffic does not need to traverse a remote VR. This effect is illustrated in Figure \ref{figure-5}.
\end{itemize}

\bigskip

\begin{figure*}[htbp]
    \centering
    \includegraphics[width=0.8\linewidth,height=0.35\textheight]{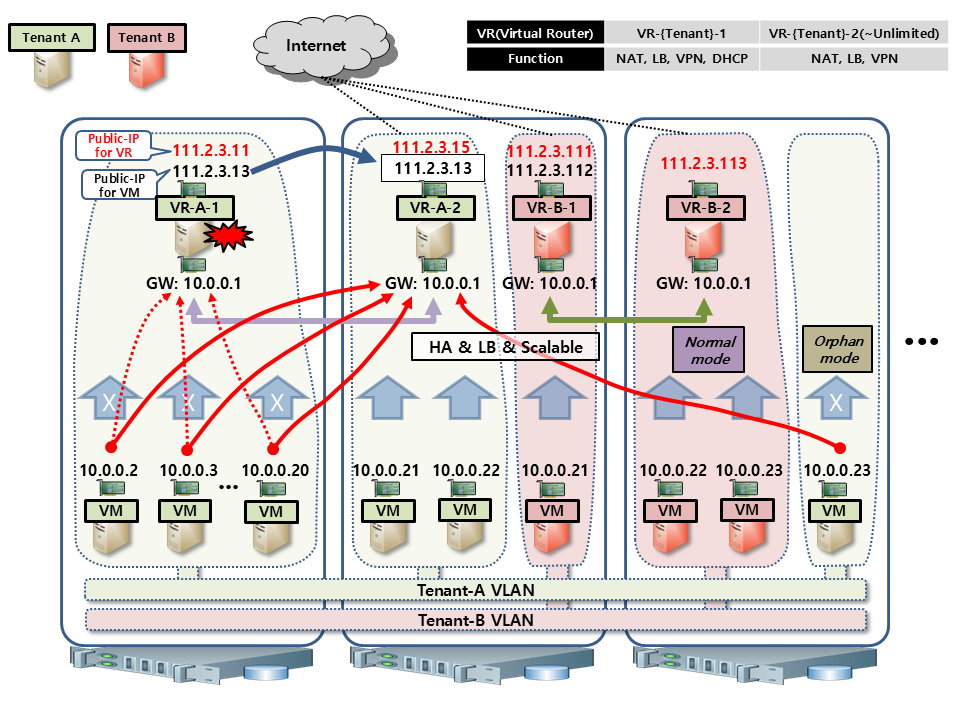}
    \caption{EYWA environments}
    \label{figure-6}
\end{figure*}

\textbf{Private Network} has traditionally been provided through VLAN (802.1Q) \cite{ref11}, which assigns a virtual LAN per tenant. However, the limited VLAN ID space restricts scalability and cannot support a very large number of tenants.

\begin{itemize}
  \item \textbf{Virtual Extensible LAN (VxLAN)} \cite{ref5}: As discussed in Section \ref{sec:background}, VLANs are constrained by the 4,094 ID limit. VxLAN overcomes this limitation by supporting up to $2^{24}$ (16,777,216) virtual networks, enabling far greater scalability. It is an overlay networking solution that effectively addresses the tenant expansion problem.
  
  \item \textbf{Large Layer-2 Network}: STP typically supports only 200–300 VMs in a single virtual LAN or subnet, and its requirement for a single active path between switches limits multipathing and network resiliency. In our MVRRP-based design, a simplified network fabric without multipathing leverages virtualization features such as fault tolerance. As a result, STP is disabled on software switches, and related issues are eliminated.  
  Switches maintain MAC tables mapping addresses to physical ports. Under MAC flooding, the table cannot locate a destination, forcing the switch to broadcast incoming traffic to all ports—behaving like a hub. In VLAN-based designs, VM MAC addresses consume limited memory in the physical switch. In contrast, VxLAN encapsulates VM MAC addresses within the host’s address space, avoiding this memory constraint.
  
  \item \textbf{Traffic Engineering}: VMs that have a local VR benefit from reduced bandwidth consumption and lower latency, since their default gateway does not require traversing a remote VR, as illustrated in Figure \ref{figure-5}.
\end{itemize}

\bigskip

\begin{figure*}[htbp]
    \centering
    \includegraphics[width=0.8\linewidth,height=0.35\textheight]{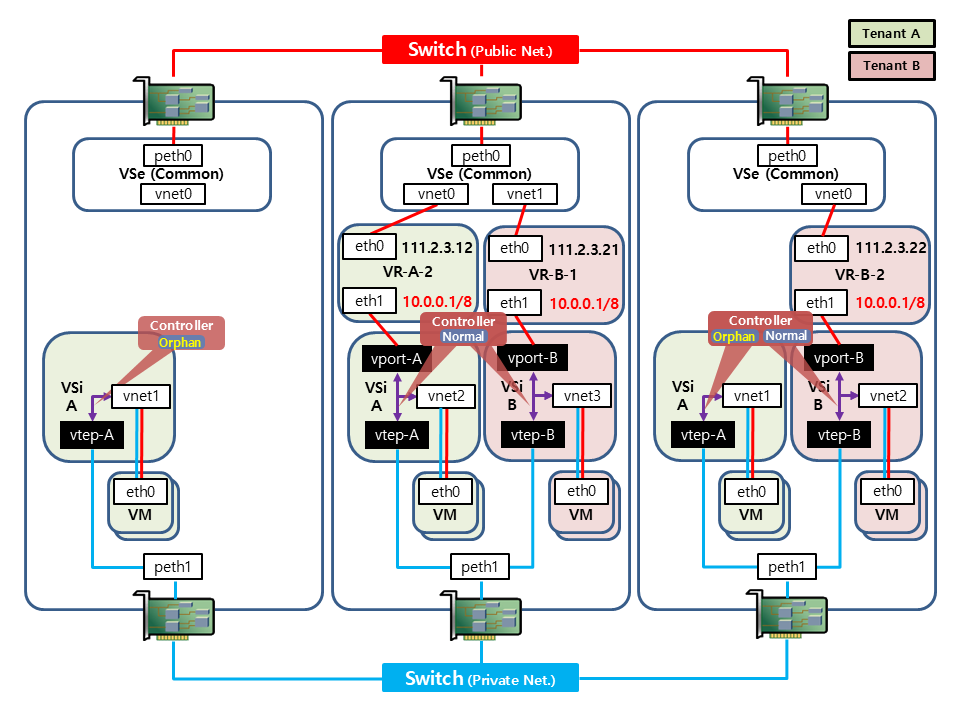}
    \caption{EYWA Agent and Mode}
    \label{figure-7}
\end{figure*}

This MVRRP and VxLAN approach still has the following limitations:

\bigskip

\textbf{Public Network} remains constrained because a maximum of only 254 VRs per tenant, each with separate private IP addresses, are available for high availability and load balancing.

\begin{itemize}
  \item \textbf{Performance}: Although 254 VRs per tenant may seem sufficient, the performance of each VR instance is relatively low compared to a physical router. As a result, the VRID limit can become a bottleneck even for medium-scale services, not just global platforms such as Google or Facebook.
  
  \item \textbf{Load Sharing}: When a master VR fails, the designated backup may already be overloaded, creating a new throughput bottleneck during takeover. In addition, the VM’s default gateway IP address, once configured, is difficult to modify during operation. Consequently, existing VM traffic cannot be redirected to newly added VRs, and a VM’s gateway cannot be updated from a remote high-latency VR to a more optimal local VR instantiated later on the same hypervisor.
  
  \item \textbf{VM Mobility}: VM migration across hypervisors is often necessary to optimize hardware utilization or reduce energy consumption. However, in MVRRP, the default gateway VR for a VM cannot always be reassigned to a more efficient local VR after migration, since changing the gateway IP during operation is not feasible.
  
  \item \textbf{Traffic Engineering}: In VRRP, master and backup VRs exchange advertisement packets at regular intervals (typically every second). These control packets consume additional network bandwidth, and shorter intervals—used to enable near real-time failover—further increase overhead.
\end{itemize}

\textbf{Private Network} issues such as the VLAN (802.1Q) limit, STP constraints, and MAC flooding have been mitigated, but some problems still remain.

\begin{itemize}
  \item \textbf{ARP Broadcast}: The MVRRP design still suffers from ARP broadcast overhead. ARP broadcasts consume network bandwidth, increase CPU utilization on every switch and server, and may introduce additional network instability. In VxLAN, ARP broadcasts are transmitted as multicast; when a reply is received, the VxLAN Tunnel End Point (VTEP) learns the MAC-to-IP mapping, and subsequent traffic is forwarded as unicast. This approach reduces bandwidth usage, but each software switch and server must still process ARP packets, which increases CPU load (Figure \ref{figure-7}). Therefore, providing a single large IP subnet remains problematic in this design.
  
  \item \textbf{Resource Consumption}: IP addresses themselves are limited resources. Each VR consumes addresses from the private IP pool (e.g., 10.0.0.1–254). In the worst case, when a tenant is assigned only a Class C subnet (10.0.0.0/24), there may be no remaining private IP addresses available for the VMs.
\end{itemize}

\bigskip

\section{\textbf{EYWA}}

The previous section showed that the MVRRP design combined with VxLAN significantly mitigated many limitations of conventional virtual networking. However, to achieve truly unlimited scalability, the remaining issues must be addressed. In this section, we propose \textbf{EYWA}, a final architecture that: (1) accommodates a very large number of tenants, (2) provides per-tenant public networking without throughput bottlenecks or single points of failure by eliminating MVRRP’s limitations in performance, availability, load balancing, VM mobility, and traffic engineering, and (3) enables each tenant to use a single large IP subnet by resolving issues related to ARP broadcasts and IP resource consumption (Figures \ref{figure-6} and \ref{figure-7}).

As in the earlier design, the VR in EYWA integrates SNAT/DNAT and layer-4 load balancing functions into a single VM instance, following the hypervisor-based service model illustrated in Figure \ref{figure-3}. For public services, EYWA also assumes the use of external DNS-based load balancing in addition to the VR’s built-in layer-4 load balancer.

A distinctive feature of EYWA is its support for two operational modes, depending on whether a VM shares the same hypervisor host with a VR:
\begin{itemize}
  \item \textbf{Normal Mode}: When a VR exists in the same internal virtual switch (VSi) as the VM, the VR is designated as the VM’s default gateway. This corresponds to a local VR configuration within the same host and tenant.
  \item \textbf{Orphan Mode}: When no VR exists in the same VSi as the VM, the VM must use a remote VR as its default gateway, even if it resides in the same tenant. This situation is illustrated in Figure \ref{figure-7}.
\end{itemize}

\bigskip

\subsection{\textbf{Agent}}

In EYWA, each VM can use multiple physically distributed VR instances that share the same private IP address (e.g., 10.0.0.1) as its default gateway, enabled by an agent running on every hypervisor host. This eliminates throughput bottlenecks and single points of failure in the public network. In addition, the agents provide a distributed proxy ARP function, allowing each tenant to maintain a single large IP subnet. Finally, EYWA supports massive multi-tenancy by leveraging VxLAN.

These capabilities cannot be achieved using only existing protocols. Instead, EYWA introduces an agent that provides functions such as VR monitoring, ARP caching, and proxy ARP. To perform these tasks, the agent continuously monitors each tenant’s Virtual Router Port (vPort) and VTEP on every hypervisor host, as illustrated in Figure \ref{figure-7}. 

Each agent operates independently, without communicating with servers or other agents.

\begin{table*}[htbp]
    \centering
    \includegraphics[width=0.9\linewidth,height=0.3\textheight]{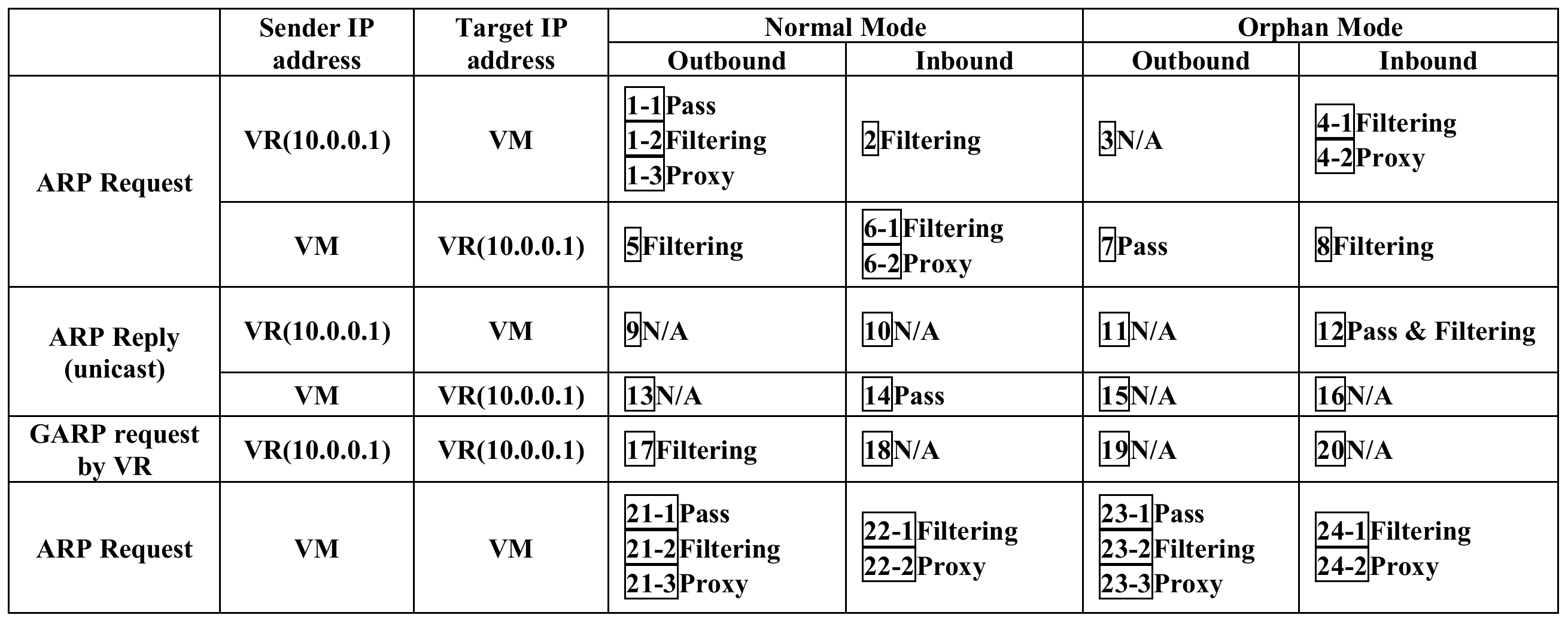}
    \caption{ARP Packet Control Rules on VTEPs}
    \label{table-1}
\end{table*}

\bigskip

\subsubsection{VR Monitoring}

EYWA’s agent monitors the vPort to assess the state and bandwidth usage of the local VR, and performs health checks through the vPort connected to the VSi. Specifically, the agent evaluates the VR’s status by monitoring ARP sessions, processing Gratuitous ARP (GARP) messages, and conducting periodic health checks. The VR’s state can also be determined passively by observing its ARP replies to VM requests. 

Based on this monitoring, the agent determines whether the system operates in \textbf{Normal Mode}—when the local VR is functioning correctly—or in \textbf{Orphan Mode}, when the local VR is unavailable. In addition, the agent collects QoS information by tracking the bandwidth utilization of the local VR through the vPort.

\bigskip

\subsubsection{ARP Caching}

The agent maintains an ARP cache to store IP-to-MAC address mappings and relies on its Proxy ARP function until the cache entries expire. For effective caching, the agent records the addresses of the local VR, local VMs, and remote VMs by monitoring ARP sessions and Gratuitous ARP (GARP) packets through the vPort and VTEP. 

To ensure consistency, the agent periodically sends ARP requests to the local VR and local VMs—though not to remote VMs—before the cache entries time out. In this way, the ARP cache is continuously refreshed and remains up to date.

\bigskip

\subsubsection{ARP Filtering \& Proxy ARP}

As described earlier, multiple VRs per tenant may share the same private IP address. To prevent conflicts, the agent filters ARP packets passing through the VTEP to ensure that VMs discover only a single gateway. This filtering avoids IP address conflicts among VRs and also enables the agent to act as a Proxy ARP, reducing ARP broadcast traffic across the VTEP. However, ARP broadcasts between local instances—such as the VR and VMs within the same VSi—are not intercepted by the agent.

In \textbf{Normal Mode}, a local VR is assigned as the VM’s default gateway. In \textbf{Orphan Mode}, the default gateway is dynamically assigned to a remote VR selected according to policy, typically based on latency and load conditions. To support these behaviors, all agents apply ARP Packet Control Rules through the VTEP, as summarized in Table \ref{table-1}. 

\bigskip

\begin{figure}[htbp]
    \centering
    \includegraphics[width=0.9\linewidth]{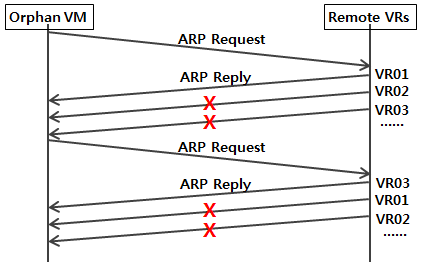}
    \caption{ARP Flux Problem}
    \label{figure-8}
\end{figure}

The detailed descriptions are as follows:

\begin{enumerate}[label=(\arabic*), align=left]
  \item \textbf{Normal Mode, Outbound ARP Request (VR $\rightarrow$ VM)}:  
  A local VR sends an ARP request to discover a local or remote orphan VM.  
  – If the target is a local VM, the packet is \framebox{1-2}Filtered, since the VM will respond directly and broadcasts must be suppressed.  
  – If the target is a remote VM, the packet is handled either by \framebox{1-3}Proxy to prevent broadcast or by \framebox{1-1}Pass, depending on whether the MAC entry exists in the agent’s cache.

  \bigskip
  \item \textbf{Normal Mode, Inbound ARP Request (VR $\rightarrow$ VM)}:  
  An incoming ARP request from a remote VR attempts to discover a remote orphan VM (via \framebox{1-1}Pass of another agent). In this case, the packet is \framebox{2}Filtered to block broadcasts and to prevent local VMs from being exposed to remote VRs.

  \bigskip
  \item \textbf{Orphan Mode, Outbound ARP Request (VR $\rightarrow$ VM)}:  
  In orphan mode, there is no local VR. Therefore, outbound ARP requests of this type are not applicable (\framebox{3}N/A).
  
  \bigskip
  \item \textbf{Orphan Mode, Inbound ARP Request (VR $\rightarrow$ VM)}:  
  A remote VR sends an ARP request to discover an orphan VM.  
  – If the target is a local VM, the agent replies on its behalf (\framebox{4-2}Proxy).  
  – Otherwise, the packet is \framebox{4-1}Filtered.

  \bigskip
  \item \textbf{Normal Mode, Outbound ARP Request (VM $\rightarrow$ VR)}:  
  A local VM sends an ARP request to discover a VR. This is \framebox{5}Filtered, since local VMs must not discover remote VRs and broadcasts must be prevented.

  \bigskip
  \item \textbf{Normal Mode, Inbound ARP Request (VM $\rightarrow$ VR)}:  
  A remote orphan VM issues an ARP request to discover a VR (via \framebox{7}Pass).  
  – If the local VR is overloaded, the packet is \framebox{6-1}Filtered for QoS.  
  – Otherwise, it is answered via \framebox{6-2}Proxy.

  \bigskip
  \item \textbf{Orphan Mode, Outbound ARP Request (VM $\rightarrow$ VR)}:  
  A local orphan VM requests to discover a remote VR. This request is \framebox{7}Passed.

  \bigskip
  \item \textbf{Orphan Mode, Inbound ARP Request (VM $\rightarrow$ VR)}:  
  A remote orphan VM sends an ARP request to discover a VR. This is \framebox{8}Filtered, because no local VR exists and broadcasts must be blocked.

  \bigskip
  \item \textbf{Normal Mode, Outbound ARP Reply (VR $\rightarrow$ VM)}:  
  A local VR replies to a remote orphan VM (\framebox{7}Pass). This is \framebox{9}N/A, since the corresponding ARP request was already processed by \framebox{6-1}Filtering or \framebox{6-2}Proxy.

  \bigskip
  \item \textbf{Normal Mode, Inbound ARP Reply (VR $\rightarrow$ VM)}:  
  A remote VR replies to a local VM. This is \framebox{10}N/A, as the ARP request was already handled by \framebox{5}Filtering.

  \bigskip
  \item \textbf{Orphan Mode, Outbound ARP Reply (VR $\rightarrow$ VM)}:  
  A local VR replying to a remote orphan VM is \framebox{11}N/A, since no local VR exists in Orphan Mode.

  \bigskip
  \item \textbf{Orphan Mode, Inbound ARP Reply (VR $\rightarrow$ VM)}:  
  Multiple remote VRs may reply to a local orphan VM (\framebox{7}Pass). This can create an ARP flux problem, where multiple replies arrive for a single request. The agent resolves this by selecting only the fastest reply and filtering the rest (\framebox{12}Pass \& Filtering), as illustrated in Figure \ref{figure-8}.

  \bigskip
  \item \textbf{Normal Mode, Outbound ARP Reply (VM $\rightarrow$ VR)}:  
  A local VM replies to a remote VR, but this is \framebox{13}N/A because the request was already handled by \framebox{2}Filtering.

  \bigskip
  \item \textbf{Normal Mode, Inbound ARP Reply (VM $\rightarrow$ VR)}:  
  A remote orphan VM replies to a local VR (via \framebox{1-1}Pass). This is allowed (\framebox{14}Pass).

  \bigskip
  \item \textbf{Orphan Mode, Outbound ARP Reply (VM $\rightarrow$ VR)}:  
  A local orphan VM replies to a remote VR (via \framebox{1-1}Pass). This is \framebox{15}N/A, since the request was already processed by \framebox{4-2}Proxy.

  \bigskip
  \item \textbf{Orphan Mode, Inbound ARP Reply (VM $\rightarrow$ VR)}:  
  A remote orphan VM replies to a local VR. This is \framebox{16}N/A, because no local VR exists in Orphan Mode.

  \bigskip
  \item \textbf{Normal Mode, Outbound GARP (VR $\rightarrow$ VR)}:  
  A local VR issues a Gratuitous ARP (GARP) to update caches and detect IP conflicts, especially when transitioning from Orphan Mode to Normal Mode. This is \framebox{17}Filtered, so only local VM caches are updated while preventing conflicts with remote VRs.  
  A GARP request is an ARP where the sender and target IPs are identical, and the destination MAC is the broadcast address (ff:ff:ff:ff:ff:ff). GARPs are used to detect conflicts, update ARP caches, and notify switches of the host’s MAC.

  \bigskip
  \item \textbf{Normal Mode, Inbound GARP (VR $\rightarrow$ VR)}:  
  A remote VR sends a GARP to update caches. This is \framebox{18}N/A, as the packet is already handled by the remote agent’s \framebox{17}Filtering.

  \bigskip
  \item \textbf{Orphan Mode, Outbound GARP (VR $\rightarrow$ VR)}:  
  A local VR would send a GARP, but this is \framebox{19}N/A since no local VR exists in Orphan Mode.

  \bigskip
  \item \textbf{Orphan Mode, Inbound GARP (VR $\rightarrow$ VR)}:  
  A remote VR sends a GARP, but this is \framebox{20}N/A as it has already been processed by the remote agent’s \framebox{17}Filtering.

  \bigskip
  \item[(21)–(24)] \textbf{Inter-VM Communication (VM $\rightarrow$ VM)}:  
  For ARP requests between VMs:
  \begin{itemize}
    \item Local VM $\rightarrow$ Remote VM: \framebox{21-3}Proxy or \framebox{21-1}Pass (\framebox{23-3}Proxy or \framebox{23-1}Pass) depending on ARP cache entries.
    \item Local VM $\rightarrow$ Local VM: \framebox{21-2}Filtering or \framebox{23-2}Filtering to prevent unnecessary broadcasts.
    \item Remote VM $\rightarrow$ Remote VM: \framebox{22-1}Filtering or \framebox{24-1}Filtering to suppress broadcasts.
    \item Remote VM $\rightarrow$ Local VM: \framebox{22-2}Proxy or \framebox{24-2}Proxy.
  \end{itemize}
\end{enumerate}

\bigskip

\subsection{\textbf{VM Migrations}}

As described in Section 4.1, EYWA addresses the limitations of MVRRP by improving performance, availability, load balancing, traffic engineering, and ARP broadcast handling through its agent design. This section highlights EYWA’s most significant advantage: live VM migration.

In conventional environments, each VM in a tenant is tied to a single VR instance configured as its default gateway. If a VM is live-migrated to another hypervisor due to VR or host overload, the default gateway IP inside the VM is difficult to change during operation. As a result, the migrated VM continues to rely on the original VR, even if a more suitable gateway is available.

\begin{figure}[tp]
    \centering
    \includegraphics[width=1.0\linewidth,height=0.23\textheight]{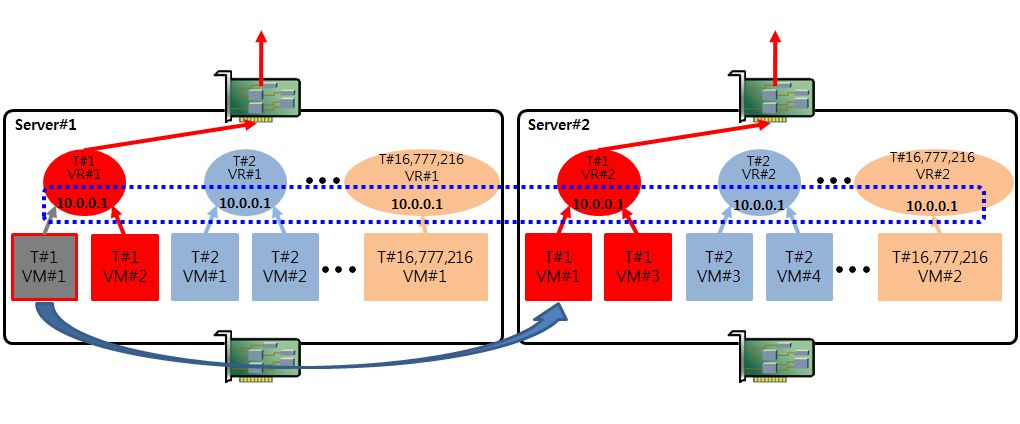}
    \caption{SNAT Traffic Flows after VM live-migration}
    \label{figure-9}
\end{figure}

In EYWA, the default gateway IP address inside the VM also remains unchanged after migration. However, the VM can seamlessly utilize a physically different and less-loaded VR as its gateway, as illustrated in Figure \ref{figure-9}. This capability reduces both network bandwidth consumption and latency. Such benefits are achieved by the EYWA agent’s packet control, which allows multiple VRs with the same private IP address to coexist within the tenant network.

\begin{figure}[bp]
    \centering
    \includegraphics[width=0.9\linewidth,height=0.2\textheight]{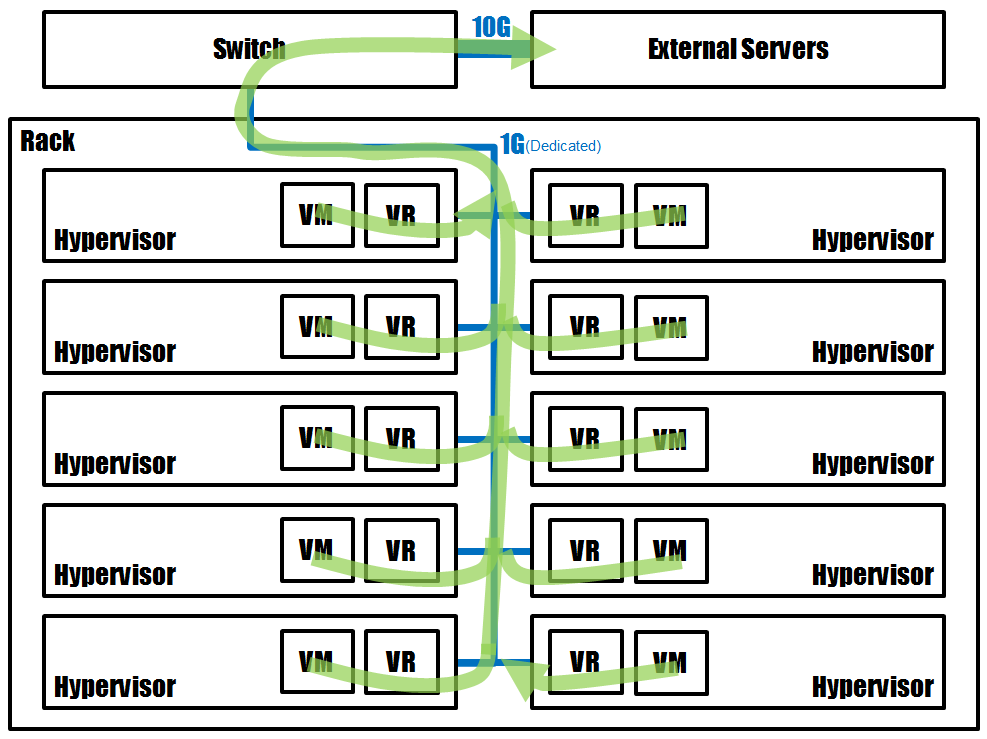}
    \caption{Test environment for north-south communication}
    \label{figure-10}
\end{figure}

\begin{figure}[tp]
    \centering
    \fbox{\includegraphics[width=1.0\linewidth,height=0.2\textheight]{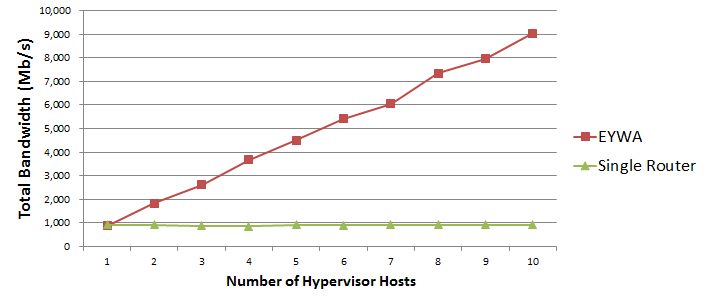}}
    \caption{Total north-south network bandwidth}
    \label{figure-11}
\end{figure}

\bigskip

\section{\textbf{Evaluation}}

We evaluate EYWA using a prototype deployed on a 10-server testbed connected through a single commodity switch (2 × 10Gbps ports and 24 × 1Gbps ports) within a single rack. Each of the 10 servers is connected via 1Gbps links. The layer-4 load balancer within each VR is implemented using the open-source HAProxy.

\bigskip

Our evaluation has two primary goals:
\begin{itemize}
  \item To demonstrate that EYWA can be constructed entirely from readily available components.
  \item To verify that our implementation resolves the public network issues discussed in Section \ref{sec:background}.
\end{itemize}

\bigskip

We exclude private network issues, such as east–west traffic within a single tenant, from this evaluation. These behaviors are well understood and are not central to our experimental objectives.

\bigskip

\subsection{\textbf{North-South Traffic}}

In this experiment, we demonstrate that all VMs can fully utilize the available physical bandwidth when communicating with external servers, without encountering throughput bottlenecks. The test environment is illustrated in Figure \ref{figure-10}.

\bigskip

\subsubsection{Outbound Communications}
\label{sec:outbound-communications}

In the baseline scenario, each hypervisor host contains both a VR and a VM belonging to the same tenant, i.e., all instances operate in Normal Mode. In this setting, every VM transmits traffic to external servers at its full physical bandwidth. As illustrated in Figure \ref{figure-11}, the aggregate outbound throughput of all VMs equals the sum of their individual link capacities. Similarly, all VMs are able to receive inbound traffic from external servers at full bandwidth, showing symmetric performance.

\bigskip

\subsubsection{Outbound Communications in the Auto-Scaling Scenario of VRs and VMs}

We further evaluate the case where the cloud platform supports auto-scaling of VRs and VMs according to predefined policies (e.g., based on network bandwidth utilization). For evaluation purposes, we emulate this behavior by manually launching or terminating VR and VM instances. The testbed environment is identical to Section \ref{sec:outbound-communications}, with the constraint that at most one VM and one VR can reside on each hypervisor host.

\begin{figure*}[htbp]
    \centering
    \fbox{\includegraphics[width=1.0\linewidth,height=0.27\textheight]{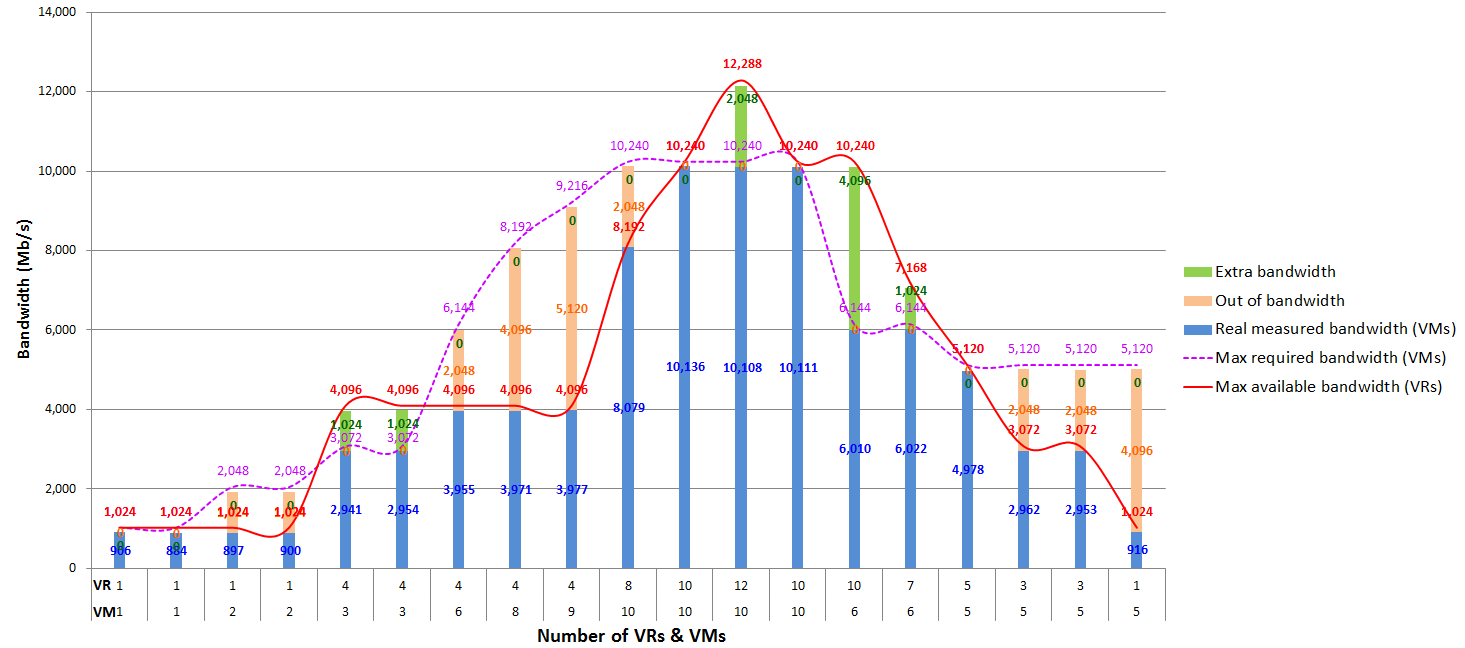}}
    \caption{Total north-south network bandwidth increased and decreased by auto-scaled VMs and VRs}
    \label{figure-12}
\end{figure*}

Figure \ref{figure-12} illustrates the aggregate outbound throughput as VRs and VMs are dynamically added or removed. The total outbound bandwidth scales proportionally with the number of active instances, and decreases accordingly when instances are terminated. The same trend is observed for inbound bandwidth.

\bigskip

\subsection{\textbf{East-West (Inter-Tenant) Traffic}}

We next evaluate the performance of inter-tenant (east–west) communications. The objective is to verify that all VMs can fully utilize the available physical bandwidth without throughput bottlenecks when communicating with VMs belonging to other tenants.

\begin{figure}[htbp]
    \centering
    \fbox{\includegraphics[width=0.95\linewidth,height=0.15\textheight]{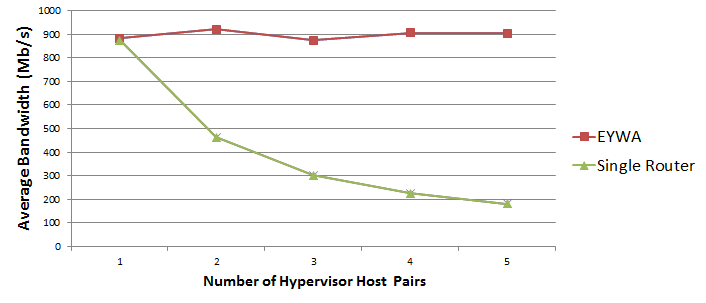}}
    \caption{Average network bandwidth per VM in inter-tenant commu-nication (1 to 1)}
    \label{figure-13}
\end{figure}

\begin{figure}[htbp]
    \centering
    \fbox{\includegraphics[width=0.95\linewidth,height=0.15\textheight]{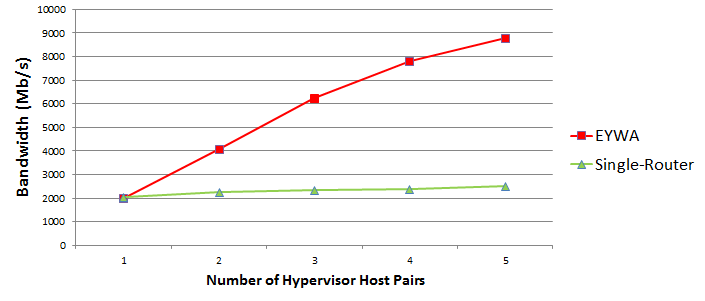}}
    \caption{Total network bandwidth of VM pairs in inter-tenant com-munication (1 to N)}
    \label{figure-14}
\end{figure}

\begin{figure}[htbp]
    \centering
    \fbox{\includegraphics[width=0.95\linewidth,height=0.15\textheight]{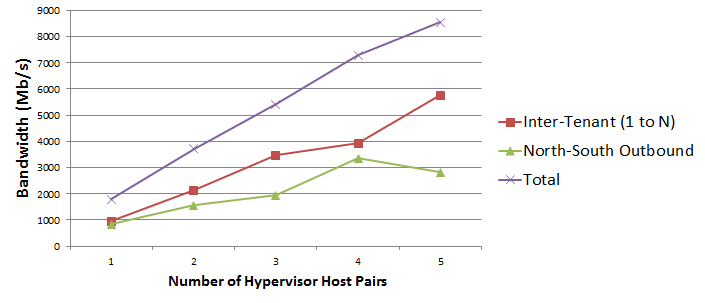}}
    \caption{Network bandwidth competition by north-south and east-west communication in EYWA}
    \label{figure-15}
\end{figure}

\bigskip

\subsubsection{1-to-1 Communications}
\label{sec:1-to-1-communications}

The test environment is identical to Section \ref{sec:outbound-communications}, except that the 10 hypervisor hosts are equally divided between tenant~A and tenant~B. Each hypervisor host contains one VR and one VM belonging to the same tenant. In this setup, every VM of tenant~A transmits traffic at full bandwidth to an idle VM of tenant~B. As illustrated in Figure \ref{figure-13}, the average outbound throughput per VM of tenant~A equals the physical link capacity of each VM.

\bigskip

\subsubsection{1-to-N Communications}

The test environment is identical to Section \ref{sec:1-to-1-communications}. Each VM of tenant~A transmits traffic at full bandwidth to all VMs of tenant~B, and symmetrically, each VM of tenant~B sends to all VMs of tenant~A. As shown in Figure \ref{figure-14}, the aggregate outbound throughput of host pairs equals the sum of the physical link capacities of all communicating VM pairs.

\bigskip

\subsection{\textbf{North-South and East-West Traffic}}

Finally, we evaluate the case where VMs simultaneously perform north–south (external) and east–west (inter-tenant) communications. The objective is to confirm that the coexistence of both traffic types does not introduce throughput bottlenecks.

\bigskip

\subsubsection{Outbound and 1-to-N Communications}

The 10 hypervisor hosts are equally divided between tenant~A and tenant~B. Each host runs one VR and two VMs: one designated for east–west communication and one for north–south communication. In this setup, each east–west VM of tenant~A transmits traffic at full bandwidth to all east–west VMs of tenant~B, while each north–south VM of tenant~A simultaneously sends traffic at full bandwidth to external servers. Tenant~B performs the same pattern of transmissions. As illustrated in Figure \ref{figure-15}, the aggregate outbound throughput of host pairs equals the sum of the physical link capacities of all VM pairs.

For comparison, in a single-router environment, the total outbound throughput of each host pair is limited to the sum of only two VMs’ physical link capacities, regardless of the number of communicating VMs.

\bigskip

\section{\textbf{Related Work}}

\textbf{Virtual Network Designs for Multi-Tenant Clouds}:  
OpenStack/Nova-Network offers three models (Flat, Flat DHCP, VLAN). The Flat and Flat DHCP models do not provide tenant-level isolation, while the VLAN model achieves isolation using VLAN tags. However, this approach suffers from pre-configuration overhead, VLAN ID limitations, a centralized Layer-3 controller bottleneck, and lacks large Layer-2 semantics.  

OpenStack/Neutron/DVR \cite{ref33} provides per-tenant virtual LANs and distributed DNAT, but retains centralized SNAT at the Network Node, and still lacks large Layer-2 semantics.  

MidoNet \cite{ref32} is an open, scalable system providing distributed networking services (switching, routing, NAT) via an agent on each hypervisor host. However, external traffic processing (routing, firewall, load balancing) is delegated to dedicated MidoNet Gateways and state databases, introducing additional infrastructure requirements.  

CloudStack \cite{ref17} offers per-tenant virtual LANs and HAProxy-based VRs, but suffers from throughput bottlenecks and SPOF, and similarly lacks large Layer-2 semantics.  

AWS VPC \cite{ref29} provides logically isolated networks using Internet Gateways that support NAT, routing, and ARP proxying. While highly available, the gateway can still become a bottleneck and cannot proxy all ARP packets in large Layer-2 topologies.  

\textbf{Physical Network Designs for Data Centers}:  
Monsoon \cite{ref1} and SEATTLE \cite{ref3} implement large Layer-2 semantics in data centers through centralized or distributed directory services. VL2 \cite{ref2} provides scalable Layer-2 semantics and hotspot-free routing with minimal host modifications. Fat-tree \cite{ref9} requires specialized routing not supported by commodity switches. TRILL \cite{ref13} and SPB \cite{ref14} standardize Layer-2 shortest path forwarding, replacing STP and enabling larger, more resilient topologies.  

\textbf{Layer-4 Load Balancing}:  
Ananta \cite{ref4} provides a distributed L4 load balancer and NAT using hypervisor agents, virtual switches, and multiplexer servers, but does not rely on DNS-based balancing. OpenStack/Neutron LBaaS \cite{ref40} allows pluggable back-end load balancers. AWS Elastic Load Balancing \cite{ref30} uses DNS-based traffic distribution across EC2 instances. Open-source solutions such as HAProxy \cite{ref35} and Linux Virtual Server (LVS) \cite{ref36} are also widely used.  

\textbf{Tunneling Protocols}:  
NVGRE \cite{ref6} uses GRE encapsulation with a 24-bit tenant ID space, supporting up to 16 million networks, similar to VxLAN. STT \cite{ref7} extends this further by introducing a 64-bit network ID space, improving scalability compared to NVGRE and VxLAN.

\bigskip

\section{\textbf{Conclusion}}

In this paper, we presented EYWA, a new virtual network architecture for multi-tenant cloud environments. EYWA accommodates a large number of tenants through isolated virtual LANs, provides per-tenant public networks without bottlenecks or single points of failure in SNAT/DNAT, and supports a single large IP subnet per tenant with scalable Layer-2 semantics. These features benefit both cloud service providers and consumers by enabling high availability, scalability, and efficient resource utilization.

EYWA is a lightweight design that can be realized using existing networking technologies without requiring modifications to host kernels, physical switches, or software switches. It also does not rely on additional centralized servers or components. Instead, it employs a distributed, independent agent deployed on every hypervisor host, eliminating centralized management overhead while ensuring scalability and resilience.  

In future work, we plan to integrate EYWA with open-source IaaS platforms and evaluate its performance in large-scale production cloud data centers. Ultimately, the scalability of virtualized environments should be constrained only by the underlying physical infrastructure, not by the virtual network design itself. EYWA advances this goal by scaling in proportion to the size of the physical network and compute cluster.

\bigskip

\section{\textbf{Acknowledgments}}

We thank the anonymous reviewers for their valuable comments, which greatly improved the final version of this paper. We are also grateful to many authors whose prior work and ideas inspired this research. Finally, we would like to acknowledge the creative inspiration provided by *Avatar* (directed by James Cameron) and Hanchul Kwon.

\end{document}